\documentclass[journal]{IEEEtran}
\usepackage{amsmath,amsfonts}
\usepackage{algorithmic}
\usepackage{algorithm}
\usepackage{array}
\usepackage[caption=false,font=normalsize,labelfont=sf,textfont=sf]{subfig}
\usepackage{textcomp}
\usepackage{stfloats}
\usepackage{url}
\usepackage{verbatim}
\usepackage{graphicx}
\usepackage{cite}
\usepackage{bm}
\usepackage{booktabs}
\usepackage{setspace}
\begin{document}

\title{Hot–Cold Tiering of HBM and High Bandwidth Flash for Agentic LLM Serving}

\author{Jongjin Baek, Won Ji, Seungjae Yoo and Joo-Young Kim
\vspace{-8mm}

\thanks{Manuscript submitted to IEEE Computer Architecture Letters, 2026.}
}

\markboth{IEEE Computer Architecture Letters,~Vol.~XX, No.~X, 2026}%
{Anonymous \MakeLowercase{\textit{et al.}}: High Bandwidth Flash and Memory for Low-Latency and Large-Scale Agentic Workloads}

\IEEEpubid{0000-0000~\copyright~2026 IEEE}

\maketitle
\begin{abstract}
Large language model (LLM) serving is increasingly agentic, with multi-turn sessions that idle between actions yet must retain their full context. 
Limited GPU memory capacity forces inactive KV states to be evicted, so resuming a session incurs either costly recomputation or slow interconnect transfers.
To address this, high bandwidth flash (HBF)—an on-package 3D-NAND memory offering orders-of-magnitude greater capacity than high bandwidth memory (HBM) at comparable read bandwidth—has emerged as a strong candidate. However, its high read energy and limited write endurance make it impractical to serve \emph{all} KV traffic.
Fortunately, our analysis shows that agentic KV states exhibit distinct access patterns: a small hot set is read for every decoding step, while a large cold pool is read only when a paused session resumes.
Exploiting this, we place the hot set in HBM and the cold pool in HBF, forming a hot–cold KV hierarchy within the GPU memory tier.
On agentic workloads with Qwen3-Coder-30B-A3B, our design delivers 14~ms time-between-tokens (TBT) and adds only $\approx\!0.1$~ms of resume latency on top of prefill, while hosting $24\times$ more concurrent sessions per GPU.
By confining HBM to the hot set, our design also cuts read power by 7.6 kW per 8-GPU node relative to serving all KV from flash—establishing HBF as a cold-tier complement to HBM rather than its replacement.
\end{abstract}
\vspace{-2mm}
\begin{IEEEkeywords}
LLM inference, KV cache, high bandwidth flash, memory hierarchy, agentic serving, hot--cold tiering.
\end{IEEEkeywords}
\vspace{-4mm}
\section{Introduction}

\begin{figure}[t]
\centering
\includegraphics[width=0.8\columnwidth]
{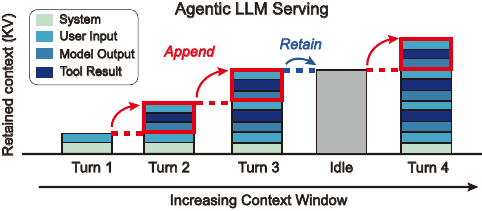}
\vspace{-3mm}
\caption{Characteristics of agentic AI}
\label{fig:agentic_ai}
\vspace{-5mm}
\end{figure}

Large Language Model (LLM) inference is becoming agentic.
A single-query session now turns into a multi-turn loop of decoding, tool calls, code execution, and user input~\cite{swebench,sarathi}.
Each turn appends the output of the model, the tool results, and the next user message.
The key-value (KV) cache then accumulates across the entire session instead of resetting per query (Fig.~\ref{fig:agentic_ai}). 
The KV cache is read at every active decoding step but remains idle and pauses during long waits for a tool or user input until it resumes~\cite{continuum, tokencake}.
Consequently, the KV cache is split into two contrasting groups. The first is a small, frequently accessed set (KV of active sessions), and the second is a large, infrequently accessed pool (KV of paused sessions), which is read only upon resumption.
Meeting tight time-between-tokens (TBT) service-level objectives (SLOs) requires keeping the active KV cache resident in HBM for fast access.
The footprint of this actively read state which grows with batch size stays small for two reasons.
First, HBM capacity is finite.
Second, under tight TBT SLOs, increasing the decode batch lowers per-request throughput~\cite{sarathi}.
The paused state behaves differently.
Because agentic workflows feature long periods of inactivity~\cite{tokencake}, the aggregate KV cache of paused sessions grows continuously over time, overflowing HBM well before a node hits its production session target~\cite{Tutti}.
To support more concurrent sessions, systems risk violating their SLOs by offloading KV states to slower memory or discarding them, leading to expensive context recomputation on resume.
This limitation is fundamentally driven by the bimodality of agentic KV cache. Paused sessions require high capacity, while both active and paused sessions require high bandwidth, and no single memory technology serves both well.
We address this bandwidth–capacity mismatch by integrating high bandwidth flash (HBF) into the GPU memory hierarchy.

HBF is an emerging memory technology that stacks 3D-NAND behind an HBM-style interface, offering greater capacity than HBM at a lower cost per bit~\cite{kyung}.
Compared to HBM, however, HBF suffers from higher latency, significantly higher per-bit read energy~\cite{haven}, and lower write endurance~\cite{H3}, which fundamentally limits write rates.
Mitigating these limitations requires integrating HBF into the memory hierarchy such that its workloads naturally mask its energy and endurance overheads.
Our contributions are as follows:
(1) identifying that KV access in agentic LLM is bimodal with a hot set read every step and a cold pool read only on resume,
(2) data movement and placement policy on a heterogeneous memory architecture exploiting this bimodality, keeping the hot set in HBM and the cold pool in HBF,
(3) trace-based simulation demonstrating that our design expands session capacity and ensures low TBT, while mitigating high HBF read energy and preserving flash endurance within warranty.
\vspace{-3mm}

\IEEEpubidadjcol

\section{Background and Motivation}

\textbf{Large Language Models}
(LLMs) generate text autoregressively in two phases. \emph{Prefill} phase processes the prompt in parallel to compute key-value (KV) tensors, and \emph{decoding} phase generates tokens sequentially by attending to past KV data.
Systems maintain a per-session \emph{KV cache} to avoid recomputation~\cite{vllm}.
Because this cache grows linearly with context length and concurrency, decoding is memory-bandwidth bound. Modern frameworks like vLLM~\cite{vllm} allocate the cache in fixed-size blocks to eliminate capacity waste.
Since autoregressive decoding only appends to the cache, a KV entry is never modified once written and is therefore immutable. We assume this setting throughout, excluding policies that rewrite KV in place such as in-session compression.

\begin{figure}[t]
\centering
\includegraphics[width=0.85\columnwidth]{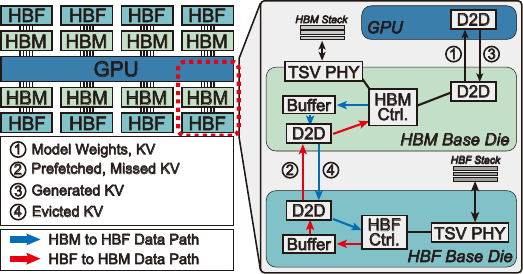}
\vspace{-3mm}
\caption{Tiered HBM+HBF architecture.
\emph{Left:} each of the memory sites of GPU  co-packages an HBF stack on an HBM stack.
\emph{Right:} Zoomed view showing the GPU connected to HBM via a die-to-die (D2D) link.
}
\vspace{-6mm}
\label{fig:Architecture}
\end{figure}

\textbf{Agentic AI Serving} wraps a model in a multi-step loop, interleaving  model calls with tool execution or user inputs that feed back into a growing  context~\cite{TheCostofDynamicReasoning}.
This creates two distinct serving dynamics, where sessions are long-lived with large KV caches, and agents spend most of their time paused and idle between actions~\cite{tokencake}.

This idle period breaks the conventional assumption that cached sessions are always actively decoding.
Because resume timing is unpredictable, the idle KV must either occupy scarce HBM the whole time or be offloaded to slower memory and fetched back on resume — trading HBM capacity against resume latency.
As a result, the KV cache divides into two populations with opposing access patterns.
A small \emph{hot set}, the KV of the currently-decoding batch, is read on every step.
Because per-request throughput declines as the batch grows~\cite{sarathi}, serving systems cap the active batch to keep this set resident in HBM.
The second population is a large \emph{cold pool}, the KV of paused sessions, which is read only on resume.
Scaling with agent concurrency, this pool quickly overflows HBM, forcing systems to provision extra hardware or suffer high eviction/recomputation penalties on resume~\cite{Tutti}.

\textbf{High bandwidth flash} 
(HBF) is an emerging memory that applies vertical die-stacking and parallel-interface design of HBM to NAND~\cite{kyung, 11573667}.
Its core structure is split into many independently addressable sub-arrays, which are read in parallel.
Together with short intra-stack data paths, die stacking enables high bandwidth.
A single stack achieves read bandwidth of about $1{,}638$~GB/s, comparable to that of an HBM4 stack~\cite{Ma_2026}.
Its capacity, however, is far larger with a single stack holding roughly $512$~GB against the $\sim\!48$--$64$~GB of HBM4~\cite{Ma_2026}.
However, HBF inherits two limitations of flash technology.
First, write endurance is limited.
Each cell tolerates a limited number of program/erase cycles. Therefore, the total writable budget per stack is the tier capacity $C$ times the rated cycle count $N_{\mathrm{PE}}$.
Dividing this budget by the effective write rate, which is the application write rate $R_{\mathrm{W}}$ times the write amplification factor $\mathrm{WAF}$, gives the expected service lifetime:
\begin{equation}
T_{\mathrm{life}} = \frac{C\,N_{\mathrm{PE}}}{R_{\mathrm{W}} \cdot \mathrm{WAF}}
\end{equation}
Because $T_{\mathrm{life}}$ scales inversely with $R_{\mathrm{W}}$, the limited endurance is only tolerable when writes are infrequent, which favors infrequently-updated data.
Second, reads are page-granular (tens of KB)~\cite{H3}, so a sub-page request still fetches and pays energy for an entire page; small or random reads therefore waste effective bandwidth and inflate per-bit read energy~\cite{haven, ju2026tilelens}.
Prior works address flash limitations through distinct strategies.
~\cite{H3} restricts storage to read-only state such as model weights and pre-computed KV, \cite{kyung} admits generated KV only when output-length predictions favor high read-to-write ratios, and \cite{pool} offloads all KV data using custom prefetching/buffering, though omitting read energy costs.
\vspace{-2mm}
\section{Design: Hot--Cold Tiering of HBM and HBF}
We hierarchically organize HBF and HBM, mapping active decoding KV caches to the bandwidth-dense, energy-efficient HBM, and the idle pool to capacity-dense HBF.
Active KV states are written to HBF only on eviction, while a resumed session is prefetched to HBM before execution resumes.

\textbf{Architecture and data paths:}
Fig.~\ref{fig:Architecture} shows the organization, which adopts the co-packaged HBM+HBF substrate proposed by~\cite{H3, pool, wang2026flashaccel}.
The HBM base die fuses the two tiers with an address router and staging buffers that double-buffer inter-tier traffic, prefetching cold blocks on resume and absorbing evictions so HBF's NAND latency hides behind the D2D transfer.
While the GPU addresses HBM directly through this base die, the HBF  is accessible exclusively via the D2D link. 
This configuration yields four distinct data paths.
\textcircled{1}~On the decode path, model weights and the active batch's KV stream from HBM into the GPU.
\textcircled{2}~When a paused session resumes, its KV is fetched from HBF back into HBM over the D2D link.
\textcircled{3}~Freshly generated KV is written back into HBM.
\textcircled{4}~Under capacity pressure, the coldest idle KV is evicted from HBM to HBF over that same link.
Paths~\textcircled{1} and~\textcircled{3} carry the hot, per-step traffic on the GPU–HBM bus. In contrast, paths~\textcircled{2} and~\textcircled{4} handle the cold-pool traffic on the HBM–HBF D2D link.
The GPU therefore reads the hot set from HBM while the base-die buffers move HBF blocks, so eviction and resume overlap with decode.

\textbf{Memory hierarchy and write policy:}
HBF can be used for the KV cache in several ways.
However, it has higher latency, lower endurance, and greater read energy than HBM.
\emph{All-KV-to-flash} methods~\cite{kyung, pool} eventually commit every KV cache to HBF, bounding endurance to the full write stream and putting HBF on the per-step write path.
We instead \emph{write-on-evict}: a KV block is written to HBF only when it is evicted from HBM, therefore a block reaches HBF only when demoted from HBM.
Because KV is immutable, a block re-admitted to HBM and evicted again does not need rewriting. The HBF copy remains valid, so \emph{write-on-evict} writes each block to HBF at most once.
This policy reduces HBF writes by confining HBF traffic strictly to eviction and resume. HBM retains model weights, active KV, and recent idle KV, while HBF acts as an overflow tier for demoted idle KV.

\begin{figure}[t]
\centering
\includegraphics[width=0.8\columnwidth]{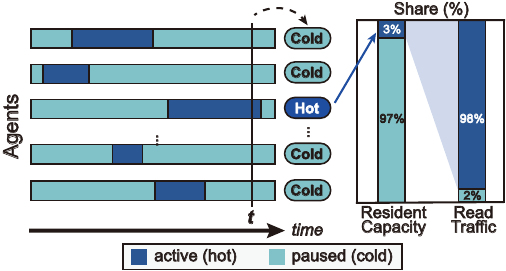}
\vspace{-3mm}
\caption{
Agentic KV access is highly bimodal. Left: an illustrative timeline where each agent alternates between active and paused. Right: measured shares at $N_a=48$, where the hot set occupies 3\% of resident capacity yet drives 98\% of read traffic.
}
\vspace{-5mm}
\label{fig:bimodal}
\end{figure}

\textbf{Eviction, resume and decode:}
When HBM fills up, the least-recently used (LRU) KV block is evicted, though any caching policy separating hot and cold sets suffices.
Active blocks are re-read every step, therefore recency naturally keeps the active set in HBM and demotes the KV states of paused sessions to HBF.
We deliberately avoid hard-pinning the KV blocks of active sessions to HBM.
Pinning may deadlock when the active set alone exceeds HBM, whereas recency gracefully degrades by spilling blocks under extreme pressure rather than failing.
On session resume, HBF-resident KV is fetched back into HBM for prefill. Sub-array parallelism amortizes flash read latency across the fetch, and it overlaps with the CPU-side processing (e.g., tokenization) that precedes each turn~\cite{chung2026characterizingcpuinducedslowdownsmultigpu}, keeping HBF off the critical path.
Following resumption and prefill, the agent enters the decoding stage, which dominates total inference time~\cite{TheCostofDynamicReasoning}.
Decode phase uses deterministic KV access patterns to prefetch required blocks one step ahead, mitigating flash-induced stalls.

\vspace{-3mm}

\section{Evaluation}
\textbf{Methodology:}
We evaluate trajectories collected with mini-swe-agent on SWE-bench Verified~\cite{swebench}, a benchmark of real software engineering tasks.
In our evaluation, we use Qwen3-Coder-30B-A3B~\cite{qwen3technicalreport}, a Mixture-of-Experts(MoE) model with 56.8~GiB of BF16 weights and 96~KiB of KV per token.
We simulate this workload in an in-house, trace-driven simulator with analytical latency and power models.
As depicted in Fig.~\ref{fig:Architecture}, the modeled node is a B200 GPU~\cite{b200} with 192~GiB of HBM3e at 8~TB/s and 2.25~PFLOP/s of compute, augmented with an HBF tier alongside its HBM.
The HBF tier is 3~TiB SLC with bandwidth matching HBM, rated at 100,000 program/erase cycles~\cite{pecycle} with 25~{\textmu}s read access latency.
As KV is immutable and appended in erase-block-aligned units, the flash avoids read-modify-write and garbage-collection rewrites. This keeps the WAF near unity ($\approx$1.02) \cite{hifc, kyung}, compared to 2 to 4 for a general-purpose SSD \cite{WAF}, leaving little room for over-provisioning to reduce it further.

To model the realistic tool execution and user idle latency, we draw each idle gap from a heavy-tailed distribution with a median of about 8 seconds and a mean of about 23 seconds, capturing tool calls that are mostly short but long-tailed~\cite{continuum} and user responses that add tens of seconds~\cite{pensieve}.
Due to scarce public agent traces, we scaled the workload to thousands of concurrent sessions to fully utilize HBF capacity by replaying measured trajectories with independent arrival phases and idle gaps.
We track each 1.5~MiB KV block as resident in HBM, HBF, or both, and apply write-back with LRU eviction. This granularity amortizes HBF's page-granular reads, whereas smaller blocks would waste sub-page reads and larger ones would coarsen eviction.
To eliminate routing-induced latency variances inherent to MoE architectures, the entire complement of expert weights is retained within HBM.
A batching scheduler~\cite{sarathi} keeps $N_a$\ sessions decoding at all times, where $N_a$\ is the number of concurrently decoding (active) agent sessions, and admits a new one whenever a pause occurs.
We sweep $N_a$ from 8 to 128, the range over which the hot set fits in HBM and the TBT SLO remains attainable.
Within each session, an agent's trajectory grows to a peak context of about 16.7k tokens, or 1.53~GiB of KV.
Across the concurrent sessions in steady state, where agents span all trajectory stages, a resident agent holds 0.82~GiB on average. 
We report this footprint, the resume hit rate (the fraction of a resuming session's KV resident in HBM), the cold-fetch and write volume over the saturated plateau of the replay, excluding its warm-up and drain~\cite{concur}.

We model read power as the sum of each memory tier's read rate multiplied by its per-bit energy.
Since HBF read energy is an unstandardized value, we treat it as a parameter and sweep it from 8 to 30~pJ/bit~\cite{haven} against HBM's $\approx$3.5~pJ/bit~\cite{rpu}.
We compare four baselines: \emph{All-KV-to-flash} method, which serves all KV from flash, CPU offload over PCIe~5.0 at 64~GB/s, or NVLink-C2C at 450~GB/s~\cite{interconnect_bw} and recomputation.

\begin{figure}[t]
\centering
\includegraphics[width=0.80\columnwidth]{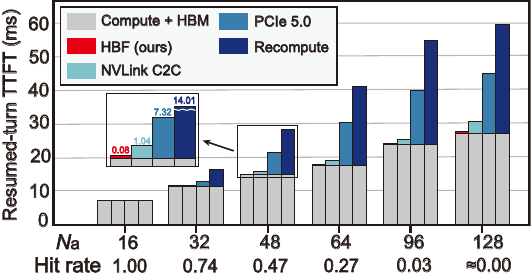}
\vspace{-3mm}
\caption{
Resumed-turn time-to-first-token (TTFT) by cold tier across $N_a$.
The grey base is the latency for prefill computation with HBM resident KV after fetching.
The colored top is the extra latency to restore the evicted blocks from each tier on resume.
The number below $N_a$ is the resume hit rate.
}
\vspace{-6mm}
\label{fig:coldtier}
\end{figure}

\begin{figure*}[t]
\centering
\includegraphics[width=0.95\textwidth]{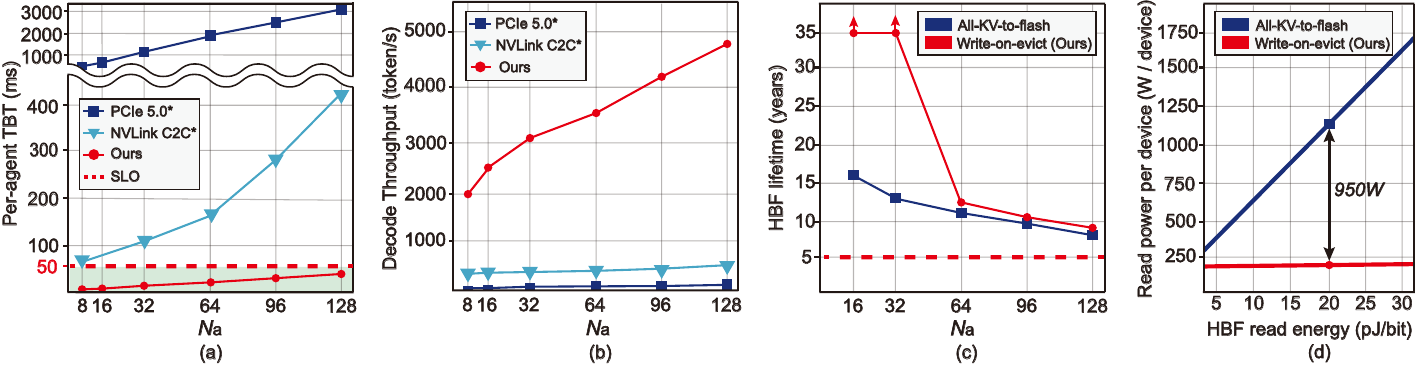}
\vspace{-3mm}
\caption{
\emph{(a)}
Per-agent TBT versus $N_a$. Asterisks(*) mark configurations without hot--cold tiering, where decode reads the hot set from the backing tier instead of HBM.
\emph{(b)}
Aggregate decode throughput for the same comparison.
\emph{(c)}
HBF lifetime versus $N_a$.
\emph{(d)}
Read power versus HBF read energy at $N_a=48$. At 20~pJ/bit, the gap is about 950~W per device.}
\vspace{-5mm}
\label{fig:integrated}
\end{figure*}

\textbf{Bimodal access:}
Fig.~\ref{fig:bimodal} confirms this bimodality on our workload.
At $N_a=48$ the hot set is about 96~GiB, fitting in HBM with headroom, yet it draws about 98\% of read traffic while occupying 3\% of KV capacity.

\textbf{Capacity and latency:}
An HBM-only device saturates at 165 sessions for the simulated trace, whereas adding a 3~TiB HBF tier expands capacity to 3,900 sessions. 
However, capacity alone is not the deciding factor, since host CPU memory offers ample capacity as well.
A critical requirement is that the memory must be fast enough to restore a session within a turn's latency budget.
Fig.~\ref{fig:coldtier} evaluates the resumed-turn TTFT by varying the cold-pool interconnect within our hot--cold framework.
As concurrency $N_a$ rises, the resume hit rate drops toward zero, increasing block misses and causing restore latencies to scale across the tiers.
At $N_a=48$, HBF adds 0.084~ms of overhead while NVLink-C2C adds 1.0~ms and PCIe adds 7~ms.
Recomputation consumes GPU compute to re-prefill evicted blocks, adding a 14~ms overhead at $N_a=48$ and over doubling turn cost at $N_a=128$.
Conversely, HBF fetches operate as pure IO off the compute path.
HBF offers capacity to hold a large cold pool at a negligible resume cost.

Since resuming a request restores its KV into HBM, the subsequent decode phase always reads KV from HBM, leaving TBT independent of the cold tier.
What matters for decode is tiering itself. Without it, the hot set may reside outside HBM and must be retrieved from the backing tier at every decoding step, failing to meet strict per-user TBT SLOs. As shown in Fig.~\ref{fig:integrated}(a), when accessing the backing tier over PCIe or NVLink, neither interconnect can sustain the bandwidth required by the hot set, leading to SLO violations.
In contrast, our design with tiering reports 14~ms per token at $N_a=48$ (72~tok/s) and 27~ms at $N_a=128$ (37~tok/s) per agent, both safely within the 50~ms SLO.
Hence, keeping the frequently read active KV on-package is crucial to meet per-step latency requirements.
Our design therefore bounds $N_a$ by how much KV HBM can hold.
Within this bound, $N_a$ trades per-agent TBT against aggregate throughput, as Fig.~\ref{fig:integrated}(b) shows throughput rising as $N_a$ scales.
The HBF tier holds the large idle pool cheaply, which suits premium latency-sensitive serving rather than throughput-maximizing batch inference.

\textbf{Endurance:}
Endurance is the main concern with HBF as a KV storage~\cite{kyung}, making lifetime a key factor.
Fig.~\ref{fig:integrated}(c) plots HBF lifetime against $N_a$\ for both write policies.
Even \emph{all-KV-to-flash}, which writes every produced KV block to HBF, clears the 5-year warranty~\cite{kyung} across the
load, falling from about 11 years at $N_a=48$ to about 8 years at $N_a=128$ as the write rate rises
with throughput. \emph{Write-on-evict}, which writes each KV block to HBF at most once on eviction and skips blocks
that never leave HBM, sits well above it. At $N_a$ below 32, nothing is
evicted and lifetime is effectively unbounded, and at $N_a=48$ it still reaches about 20 years. The
two curves converge at high $N_a$, where the working set outgrows HBM and \emph{write-on-evict} must evict
nearly as often as \emph{all-KV-to-flash}.
Endurance is therefore not a barrier to placing the cold pool in
HBF but an axis our design improves.

\textbf{Read energy:}
Serving all KV from flash~\cite{kyung, pool} pays HBF's read energy on every actively read KV, whereas our split leaves HBF carrying only the cold pool, read once per resume.
Across the load, the read rate rises from 6~GB/s at $N_a=16$ to 38~GB/s at $N_a=128$.
Fig.~\ref{fig:integrated}(d) highlights the power difference between \emph{all-KV-to-flash} and \emph{write-on-evict} policies. Assuming an HBF energy of 20~pJ/bit based on the power budget in~\cite{H3}, a \emph{write-on-evict} policy saves about 950~W per device, or 7.6~kW across an 8-GPU node.
The difference scales with the energy gap between HBM and HBF, from about 2.1 to 12.2~kW per node across the swept read-energy range.
This result shows that sourcing hot KV from HBF rather than HBM is impractical from a power consumption perspective.

\vspace{-3mm}
\section{Conclusion}

Prior HBF designs either place read-only KV in HBF~\cite{H3} or place all KV in HBF~\cite{kyung, pool}.
We instead match the memory hierarchy to the bimodal access pattern of agentic LLM KV. The hot set read every step stays in HBM while only the cold idle pool moves to capacity-optimized HBF. This supports an order of magnitude more concurrent sessions per GPU and cuts memory-read power by about 7.6~kW per 8-GPU node versus \emph{all-KV-to-flash}, with decoding running entirely on HBM.
HBF thus serves as an efficient cold tier in a hot--cold KV hierarchy, not a replacement for HBM.

\vspace{-3mm}

\bibliographystyle{IEEEtran}
\begin{spacing}{0.85}
\bibliography{refs}   

@inproceedings{vllm,
  title     = {Efficient Memory Management for Large Language Model Serving with {PagedAttention}},
  author    = {W. Kwon and others},
  booktitle = {SOSP},
  year      = {2023}
}

@article{kyung,
  title   = {High-Bandwidth Flash for {KV} Caches: Endurance and Performance Implications},
  author  = {K. Kyung and others},
  journal = {IEEE CAL},
  year    = {2026},
  doi     = {10.1109/LCA.2026.3695938}
}

@article{haven,
      title={HAVEN: High-Bandwidth Flash Augmented Vector Engine for Large-Scale Approximate Nearest-Neighbor Search Acceleration}, 
      author={P. Hsu and others},
      year={2026},
      archivePrefix={arXiv},
      primaryClass={cs.AR},
}

@misc{swebench,
  title     = {{SWE-bench-Verified and mini-swe-agent}},
  author    = {{SWE-bench Team}},
  note = {\url{https://www.swebench.com/verified.html}},
  year      = {2025}
}

@inproceedings{sarathi,
  title     = {Taming Throughput-Latency Tradeoff in {LLM} Inference with {Sarathi-Serve}},
  author    = {A. Agrawal and others},
  booktitle = {OSDI},
  year      = {2024}
}

@inproceedings{pensieve,
  title     = {Stateful Large Language Model Serving with {Pensieve}},
  author    = {L. Yu and others},
  booktitle = {EuroSys},
  year      = {2025}
}

@article{continuum,
  title   = {Continuum: Efficient and Robust Multi-Turn {LLM} Agent Scheduling with {KV} Cache Time-to-Live},
  author  = {H. Li and others},
  journal = {arXiv},
  year    = {2026}
}

@article{chung2026characterizingcpuinducedslowdownsmultigpu,
  title   = {Characterizing {CPU}-Induced Slowdowns in Multi-{GPU} {LLM} Inference},
  author  = {E. Chung and others},
  journal = {arXiv},
  year    = {2026}
}

@article{tokencake,
title={TokenCake: A KV-Cache-centric Serving Framework for LLM-based Multi-Agent Applications}, 
author={Zhuohang Bian and others},
journal = {arXiv},
year={2026},
}

@article{Tutti,
      title={Tutti: Making SSD-Backed KV Cache Practical for Long-Context LLM Serving}, 
      author={Shi Qiu and others},
      year={2026},
      eprint={2605.03375},
      archivePrefix={arXiv},
      journal = {arXiv},
      primaryClass={cs.OS},
}

@article{H3,
  author={Ha, Minho and others},
  journal={IEEE CAL}, 
  title={H3: Hybrid Architecture Using High Bandwidth Memory and High Bandwidth Flash for Cost-Efficient LLM Inference}, 
  year={2026},
  volume={},
  number={},
  pages={},
  doi={10.1109/LCA.2026.3660969}
}

@inproceedings{
hifc,
title={Hi{FC}: High-efficiency Flash-based {KV} Cache Swapping for Scaling {LLM} Inference},
author={I. Jeong and others},
booktitle={NeurIPS},
year={2026},
}

@article{
b200,
title={NVIDIA DGX B200},
author={Nvidia},
booktitle={},
year={2025},
note = {\url{https://resources.nvidia.com/en-us-dgx-systems/dgx-b200-datasheet?ncid=no-ncid}}
}

@INPROCEEDINGS{TheCostofDynamicReasoning,
  author={K. Jiin and others},
  booktitle={HPCA}, 
  title={The Cost of Dynamic Reasoning: Demystifying AI Agents and Test-Time Scaling from an AI Infrastructure Perspective}, 
  year={2026},
  volume={},
  number={},
  pages={},
  doi={10.1109/HPCA68181.2026.11408569}
}

@article{concur,
      title={CONCUR: High-Throughput Agentic Batch Inference of LLM via Congestion-Based Concurrency Control}, 
      author={Qiaoling Chen and others},
      year={2026},
      archivePrefix={arXiv},
      primaryClass={cs.DC},
}

@article{
WAF,
title={Write Amplification Factor for general SSDs},
author={Micron},
year={2025},
url = {https://www.crucial.com/support/articles-faq-ssd/why-does-ssd-seem-to-be-wearing-prematurely}
}

@inproceedings{
pecycle,
title={P/E Cycle Endurance for SLC NAND},
author={Micron},
booktitle={},
year={2026},
note = {\url{https://www.micron.com/products/storage/nand-flash}}
}

@INPROCEEDINGS{rpu,
  author={M. Adiletta and others},
  booktitle={HPCA}, 
  title={RPU – A Reasoning Processing Unit}, 
  year={2026},
  volume={},
  number={},
  pages={},
  doi={10.1109/HPCA68181.2026.11408490}}

@article{Ma_2026,
   title={Challenges and Research Directions for Large Language Model Inference Hardware},
   volume={59},
   ISSN={1558-0814},
   DOI={10.1109/mc.2026.3652916},
   number={},
   journal={Computer},
   publisher={Institute of Electrical and Electronics Engineers (IEEE)},
   author={Ma, Xiaoyu and Patterson, David},
   year={2026},
}

@article{qwen3technicalreport,
      title={Qwen3 Technical Report}, 
      author={An Yang and others},
      year={2025},
      eprint={2505.09388},
      archivePrefix={arXiv},
      journal={arXiv},
      primaryClass={cs.CL},
}

@inproceedings{
interconnect_bw,
title={NVIDIA Grace Hopper Superchip Architecture In-Depth},
author={Nvidia},
booktitle={},
year={2022},
note = {\url{https://developer.nvidia.com/blog/nvidia-grace-hopper-superchip-architecture-in-depth/}}

}

@ARTICLE{pool,
author={J. Park and others},
journal={ IEEE CAL },
title={{ HBM-HBF-Centric Memory Pooling Architecture With Custom Base Die for Terabyte-Scale LLM Inference }},
volume={},
number={},
ISSN={1556-6064},
year={2026},
doi={10.1109/LCA.2026.3703982},
publisher={IEEE Computer Society},
address={Los Alamitos, CA, USA},}

@ARTICLE{11573667,
 author={Son, Dowon and others},
 journal={IEEE Computer Architecture Letters},
 title={Exploring High-Bandwidth Flash for Modern LLM Inference: Opportunities and Challenges},
 year={2026},
 volume={25},
 number={2},
 pages={251-254},
 doi={10.1109/LCA.2026.3705817}}

@article{ju2026tilelens,
 title={{TileLens}: Efficiently Using Large-Granularity Memory Systems with Transparent Two-Dimensional Memory Layout},
 author={Ju, Jae Hyung and others},
 journal={arXiv preprint arXiv:2607.04031},
 year={2026}
}

@article{wang2026flashaccel,
 title={{FlashAccel}: Leveraging High-Bandwidth Flash for High-Throughput LLM Inference},
 author={Wang, Xinyu and others},
 journal={arXiv preprint arXiv:2607.10186},
 year={2026}
}
\end{spacing}

\end{document}